\documentclass[twocolumn,prb,superscriptaddress,showpacs,preprintnumbers]{revtex4}

\usepackage{graphicx}
\usepackage{amsmath,graphicx,dcolumn}
\usepackage[usenames]{color}

\begin{document}

\title{Quantifying structural damage from self-irradiation in a plutonium superconductor}

\author{C. H. Booth}
\affiliation{Chemical Sciences Division,
Lawrence Berkeley National Laboratory,
Berkeley, California 94720, USA}
\author{E. D. Bauer}
\affiliation{Materials Physics and Applications Division, Los Alamos National Laboratory,
Los Alamos, New Mexico 87545, USA}
\author{M. Daniel}
\affiliation{Chemical Sciences Division,
Lawrence Berkeley National Laboratory,
Berkeley, California 94720, USA}
\author{R. E. Wilson}
\altaffiliation[Permanent address: ]{Chemistry Division, Argonne National Laboratory, 9700 South Cass Avenue, Argonne, IL 60439-4831, USA}
\affiliation{Chemical Sciences Division,
Lawrence Berkeley National Laboratory,
Berkeley, California 94720, USA}
\author{J.~N.~Mitchell}
\affiliation{Materials Science and Technology Division, Los Alamos National Laboratory,
Los Alamos, New Mexico 87545, USA}
\author{L. A. Morales}
\affiliation{Materials Science and Technology Division, Los Alamos National Laboratory,
Los Alamos, New Mexico 87545, USA}
\author{J. L. Sarrao}
\affiliation{Materials Physics and Applications Division, Los Alamos National Laboratory,
Los Alamos, New Mexico 87545, USA}
\author{P. G. Allen}
\affiliation{Materials Science and Technology Division, Lawrence Livermore
National Laboratory, Livermore, California 94550, USA}

\date{\today}
%\date{\today, \xxivtime}

%\preprint{LBNL-61766}

\begin{abstract}
The 18.5 K superconductor PuCoGa$_5$ has many unusual properties, including
those due to damage induced by self-irradiation. The superconducting 
transition temperature decreases sharply with time,
suggesting a radiation-induced Frenkel defect concentration 
much larger than predicted by current radiation damage theories.  Extended 
x-ray absorption fine-structure measurements demonstrate that while the local 
crystal structure in fresh material is well ordered, aged material is 
disordered much more strongly than expected from simple defects, consistent 
with strong disorder throughout the damage cascade region.  These data 
highlight the potential impact of local lattice distortions relative to defects 
on the properties of irradiated materials and underscore the need for more 
atomic-resolution structural comparisons between radiation damage experiments 
and theory.

\end{abstract}

\pacs{71.27.+a, 74.70.Tx, 61.80.-x, 61.10.Ht}
%EXAFS and XANES in condensed matter, 61.10.Ht
%Structure irradiation effects on, 61.80.–x
%Strongly correlated electron systems, 71.27.+a
%Heavy-fermion solids superconductivity, 74.70.Tx

\maketitle

%*****************************************************************************
%*****************************************************************************
\section{Introduction}
Plutonium is arguably the most complex and least understood of all 
elements, ultimately due to the propensity of
its 5$f$-electrons to simultaneously reside in bonding and non-bonding 
electronic states. Theoretical models of $\delta$- and $\alpha$-Pu, for instance, show 
promise for explaining this complex 
behavior,\cite{Shim07,Dai03,Savrasov01} but assume a homogeneous crystalline structure despite 
unavoidable self-irradiation damage, which can significantly alter magnetic and
electronic properties. Superconductors 
provide a path for elucidating radiation damage effects\cite{Sweedler79} since their 
properties are especially vulnerable 
to atomic-level structural disorder.  Here, we report extended x-ray absorption 
fine-structure (EXAFS) measurements on the PuCoGa$_5$ superconductor\cite{Sarrao02} that 
demonstrate the local structure of aged 
material is damaged at least an order of magnitude faster than theoretical predictions focusing on Frenkel defects indicate.\cite{Wolfer00}
These results explain the sharp reduction of the superconducting critical temperature, $T_c$, with time
and underscore the need for improved radiation damage models 
relevant not only for understanding plutonium superconductors,
but also for Pu metal and other radioactive materials.

In addition to changing fundamental properties in elemental Pu and PuCoGa$_5$, radiation damage 
affects many aspects of science and industry, for example, in semiconductors, nuclear power 
generation and its associated waste disposal, and the aging nuclear stockpile. Consequently, 
radiation damage has been studied for over a century, and intensely since World War II. Although 
experimental measures have provided ample verification of long-range ($>$10 nm) structural effects 
due to radiation damage, atomic-resolution descriptions of the damage induced in a crystal have 
relied almost exclusively on theoretical calculations. This reliance has been predictably dangerous. 
Recent\cite{Farnan07,Farnan01} nuclear magnetic resonance (NMR) experiments reveal that radiation 
damage accumulates about 5 times faster in zircons and other ceramics than calculations indicate, 
calling into question both the theoretical calculations and the viability of  nuclear waste 
containment schemes. Experimental studies have not generally kept pace 
with theoretical treatments of fully relaxed damage cascade structures at the atomic level. Damage 
cascade calculations have not been quantitatively compared to 
any atomic-resolution experiments until very recently,\cite{Farnan07,Farnan01} and 
never to our knowledge in intermetallics\cite{AverbackNote} or technologically 
important materials such as $\delta$-Pu. 
Measurements of such structural 
changes are vital to verify and improve these theories, and for comparison to more complex theories of dislocation loops, He bubble formation, and volume expansion. 

Because large regions of a given sample remain in an undamaged state while damaged regions no longer have translational symmetry, traditional scattering techniques such as Rietveld analysis of powder diffraction data have not generated a quantitative measure of the damage fraction of a material. Local probes that treat ordered and disordered regions on an equal footing should be able to provide a more detailed, atomic-level description of the damage. NMR is one such technique, although it does not easily lend itself to a direct structural interpretation. The EXAFS technique employed here provides another local probe and has the advantage that it gives radial pair-distance distribution information around a specific atomic species, since it relies on the backscattering of a photoelectron from a core excitation.

\begin{figure}
\includegraphics[width=3.5in,trim=0 0 0 0]{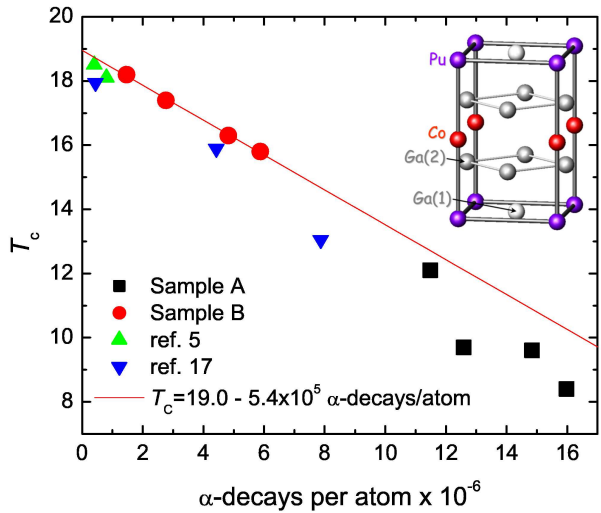}
\caption{(Color online) 
The superconducting transition temperature ($T_c$) as measured by magnetometry 
as a function of the $\alpha$-decays per atom for this study (samples {\bf A} 
and {\bf B}) and in the literature 
(1 yr=4.9$\times10^{-6}$~$\alpha$-decays per atom for these samples). 
A linear decay 
of $T_c$ (\textcolor{red}{---}) starting from about 19.0 K and decaying at a 
rate of about 5.4$\times10^{5}$ K per $\alpha$-decay per atom is also shown for 
reference.
The inset shows the crystal structure of PuCoGa$_5$.\cite{Sarrao02}
}
\label{Tc_fig}
\end{figure}

To investigate structural radiation damage effects at the atomic level in 
plutonium, we employ superconducting PuCoGa$_5$ (Fig. \ref{Tc_fig}).  The 
unusual properties of this 18.5 K superconductor, i.e., a nearly ten-fold 
higher $T_c$ than any other heavy-fermion related intermetallic and an 
electronic structure that bears a strong resemblance to $\delta$-Pu, have been 
well characterized.\cite{Sarrao02,Joyce03,Curro05} Superconducting properties 
that change with time due to plutonium  $\alpha$-decay include a  decrease in 
$T_c$ by $\sim$0.2 K per month (Fig. \ref{Tc_fig}),   the upper critical 
field $H_{c2}\approx 70$ T, and large critical current density 
$J_c > 10^4$ A/cm$^2$ for $T>0.9 T_c$,\cite{Sarrao02} likely caused by 
damage-induced scattering and pinning centers. In addition, the impurity 
scattering rate inferred from NMR experiments follows the observed change in 
$T_c$.\cite{Curro05}

%*****************************************************************************
%*****************************************************************************

Although a modern theoretical 
treatment of radiation damage in PuCoGa$_5$ is currently not available, we 
utilize calculations on the structurally similar $\delta$-Pu system for 
comparison.\cite{Sarrao02}  In these models, the 
$\alpha$-particle generated by the decay of a $^{239}$Pu nucleus has about 
5 MeV of energy and ballistically generates $\sim$300 Frenkel defect pairs over
a distance of nearly a micron.\cite{Wolfer00} Most of the damage,
however, is done by the recoiling $^{235}$U nucleus with 86 keV, which 
produces $\sim$2300 Frenkel pairs. A typical generated damage cascade extends over 
nearly 10 nm, with a defect volume fraction of about 3\%.
It is important to note that only the effect of Frenkel-type defects are 
considered in this view of radiation damage; however, these
models form the basis of most modern theories such as molecular dynamics 
theories (for example, see Ref. \onlinecite{Valone04}). In fact, lattice 
relaxation can occur during the intermediate time-scales after these 
displacement events, and have been explored in various metallic systems with 
molecular dynamics and kinetic Monte Carlo techniques. These latter 
calculations show that the effective number of defects is reduced by as much 
as an order of magnitude within only a few picoseconds due to additional 
defect migration.\cite{Rubia99} These values should not change substantially 
in PuCoGa$_5$, and in fact a rough TRIM code\cite{TRIM} calculation (which 
does not include the longer-time scale defect migration) using default values 
generates similar damage rates and ranges to the models above.
We therefore expect far fewer than the initial $N_\textrm{D}\sim$2600 pair defects to survive the defect migration per 
$\alpha$-decay.  Using an $\alpha$-decay rate $\lambda_\alpha\approx3.4\times10^{-5}$ per Pu per year
from the samples discussed below, the expected upper-limit damage fraction  
$f_\textrm{tot}=2 N_\textrm{D}\lambda_\alpha/7\approx2.5\%$ after one year, 
with 1/7$^\textit{th}$ of the atoms being Pu. $T_c$ is reduced by 50\% after
about 3 years (Fig. \ref{Tc_fig}).
These damage estimates therefore indicate that, including the expected 
additional defect migration, $f_\textrm{tot}(T_c/T_{c0}=50\%) \ll 7.5$\%. This 
estimate is inconsistent with what one expects in a short coherence length 
superconductor: 
Within the Bogoliubov-de Gennes (BdG) formalism with strong scattering and a 
coherence length $\xi_0\approx$ 2.1 nm\cite{Sarrao02}), the {\it lower limit} 
(strong scattering limit) damage fraction is
$f_\textrm{BdG}(T_c/T_{c0}=50\%) \approx$15\%.\cite{Franz97}  This value is consistent with recent 
studies of Ce$_{1-x}$La$_x$CoIn$_5$\cite{Petrovic02} and 
CeCoIn$_{5-x}$Sn$_x$\cite{Daniel05b} where the substitutions are mostly within 
the superconducting planes, and therefore likely produce strong scattering. 
The critical damage fraction should be higher when defects are randomly 
distributed, as with radiation damage.

This paper continues with a description of the sample characterization and other experimental methods (Sec. \ref{Methods}), details of the data analysis and results (Sec. \ref{Results}), a discussion of the implications of these results (Sec. \ref{Discussion}) and a concluding summary (Sec. \ref{Conclusion}).

%*****************************************************************************
%*****************************************************************************
\section{Experimental Methods}
\label{Methods}

Two PuCoGa$_5$ samples were synthesized from Ga flux.\cite{Sarrao02} At the time of the 
most recent x-ray measurements, one was about 3 years old (sample {\bf A}), and the 
other was about 1 year
old (sample {\bf B}). Superconducting critical temperatures were measured in a Quantum Design Magnetic Properties Measurement System as the point at which diamagnetism was observed in 10 Oe (Fig. \ref{Tc_fig}). The isotopic content 
is the same for both samples with the main radioactivity coming from 
93.93\% $^{239}$Pu,
5.85\% $^{240}$Pu, and
0.12\% $^{241}$Pu. 
The accumulated dose has been shown to be a reasonable indicator
of how sample properties change with time,\cite{Jutier05} and
we report the sample age in units of $\alpha$-decays per atom (1 yr=4.9$\times10^{-6}$~$\alpha$-decays per atom in the formula unit).  We do not 
use the more common ``displacements per atom'' (dpa) unit, because it
generally assumes the number of displaced atoms (Frenkel pairs) is known
from theoretical models, and we show below that this assumption 
may not be correct.

%*****************************************************************************
\begin{figure}
\includegraphics[width=3.5in]{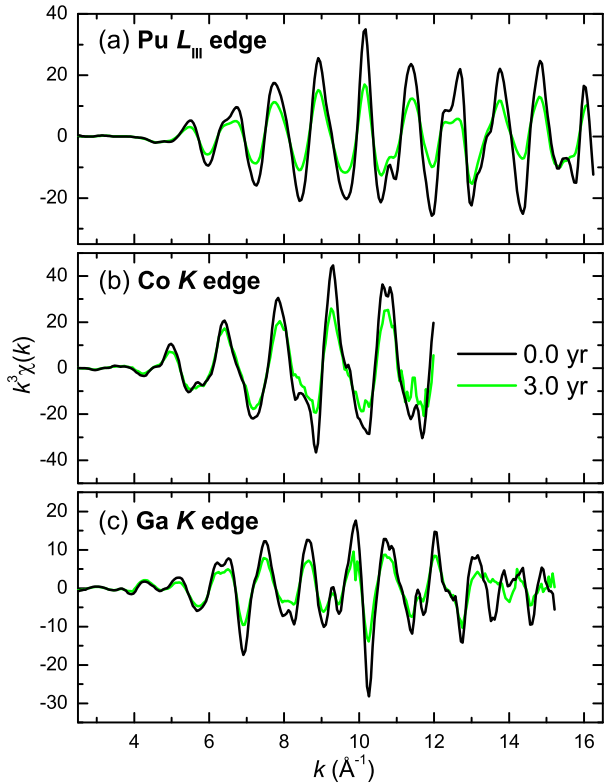}
\caption{(Color online) 
EXAFS data $k^3\chi(k)$ vs. $k$ from the (a) Pu $L_\textrm{III}$ edge, (b) the Co $K$ edge, and (c) the Ga $K$ edge, for a fresh ({\bf ---}, $0.2\times10^{-6}$~$\alpha$-decays per atom) and an aged ({\bf \textcolor{green}{---}}, 14.5$\times10^{-6}$~$\alpha$-decays per atom) sample.}
\label{ks_fig}
\end{figure}
%*****************************************************************************

%*****************************************************************************
\begin{figure}
\includegraphics[width=3.5in]{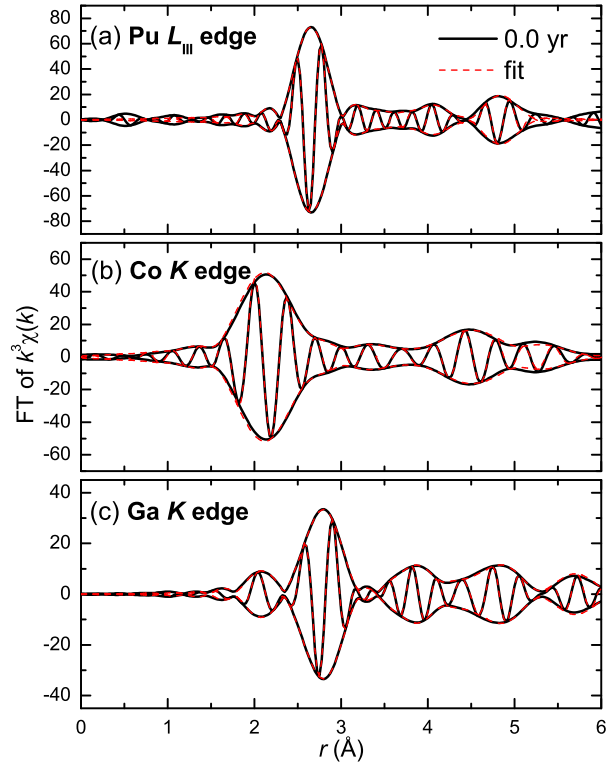}
\caption{(Color online)
Fourier transform (FT) of the $k^3\chi(k)$ data of the fresh sample ({\bf ---}) in Fig. \ref{ks_fig}, together with a fit to these data (\textcolor{red}{{\bf -~-~-}}) .  The fit quality is such that the fit is difficult to distinguish from 
the data. The outer envelope
is $\pm$ the transform amplitude and the inner modulating line is the real part of
the complex transform. The Pu edge data (a) are transformed between
2.5-16.0 \AA$^{-1}$, Gaussian broadened by 0.3 \AA$^{-1}$, and are fit between 2.0 and 5.0 \AA{}. The Co edge data (a) are transformed between
2.5-11.0 \AA$^{-1}$, Gaussian broadened by 0.3 \AA$^{-1}$, and are fit between 2.0 and 5.5 \AA{}. The Ga edge data (a) are transformed between
2.5-15.0 \AA$^{-1}$, Gaussian broadened by 0.3 \AA$^{-1}$, and are fit between 1.6 and 5.0 \AA{}. }
\label{rs_fig}
\end{figure}
%*****************************************************************************

The material was triply contained for EXAFS experiments
using epoxy- and indium-sealed kapton windows, and placed into a LHe flow
cryostat at $T\leq$30 K. Sample {\bf A} was initially measured (measurements with $\leq 4 \times 10^{-6}$~$\alpha$-decays per atom) as a single crystal in fluorescence mode using 30-element Ge detectors, with the data corrected for dead time and self absorption.\cite{Booth05a} Otherwise the samples were ground for the EXAFS experiments and passed through a
32 $\mu$m sieve, with about 8 mg of this powder mixed with dried boron
nitride and packed into an aluminum frame. Transmission and fluorescence data agree quantitatively. This sample mass resulted in a change in absorption
across the Pu $L_\textrm{III}$ edge $\Delta \mu_a$ of  $\sim$0.5 absorption
lengths, whereas $\Delta \mu_a \sim$0.3 and $\Delta \mu_a \sim$2.2  absorption
lengths across the Co and Ga $K$ edges, respectively. 
EXAFS spectra were collected at the
Stanford Synchrotron
Radiation Laboratory on beamlines 10-2 and 11-2 over a three year period, at
the Pu $L_\textrm{III}$, Co $K$, and Ga $K$ edges, generally using a
half-tuned, double crystal Si(220) monochromator. The monochromator resolution
was adjusted such that it was well below the
core-hole lifetime at a given edge.
The data were analyzed using standard
procedures.\cite{Li95b} In particular, the embedded atom absorption $\mu_0$
was determined using a cubic spline with between 4 and 6 knots over the data
range, which was typically about 1 keV above the absorption threshold, $E_0$, as determined by the position of the half-height of the absorption change at the edge.
The data were fit in $r$-space using the RSXAP package\cite{RSXAP} with theoretical scattering functions generated by FEFF7.\cite{FEFF7}

%*****************************************************************************
\begin{table*}
\caption{
EXAFS fit results for the Pu $L_\textrm{III}$ and the Co $K$ edges on a fresh (0.2$\times10^{-6}$~$\alpha$-decays per atom) sample of PuCoGa$_5$. Fit and transform ranges are listed in Fig. \ref{rs_fig}. All single-scattering peaks
within the fit range are included in these fits. Multiple scattering was only included in the Ga edge fits to avoid errors originating from peak overlap, but the results from these scattering paths are in themselves unreliable and so are not reported.  Coordination numbers, $N$, are held fixed to the nominal structure. $S_0^2$, $\Delta E_0$, and $R(\%)$ are 0.89(5), -4.7(3) eV
and 3.6\% for the Pu edge, 0.85(5), 5(1) eV, and 6.1\% for the Co edge, and 0.85(3), -1.7(4), and 6.9\% for the Ga edge,
respectively. The number of free parameters in the fits are 10, 14, and 15 for the Pu, Co , and Ga edges, respectively, and are far below the number of independent data points as given by Stern's rule.\cite{Stern93} The quoted error on each quantity is the greater of that obtained by comparisons
to standard materials,\cite{Li95b} repeated measurements, and a Monte-Carlo method.\cite{Lawrence01} The nature of the Ga local environment required constraints on several parameters to obtain meaningful fits, as indicated. Constraints on the pair distances in the Ga edge fits assume a tetragonal 115 structure such that Ga(1) sits at the center of the Pu $ab$ face and a plane of Co atoms splits a plane of Ga(2).
}
\begin{ruledtabular}
\begin{tabular}{lcccccc}
& $N$ & $\sigma^{2}$ (\AA$^2$) & $R$ (\AA) & $R_\textrm{diff}$\cite{Sarrao02} (\AA) &
$\Theta_\textrm{cD}$ (K) & $\sigma_\textrm{stat}^2$ (\AA$^2$) \\
\colrule
Pu-Ga(1)/Ga(2) & 12
& 0.00174(9) &  2.97(1) & 2.993 & 330(20) & -0.0001(2) \\
Pu-Co & 2
&  0.0019(4) &  3.38(2) & 3.393 & 420(30) & 0.0002(2)\\
Pu-Pu & 4
&  0.0016(2) &  4.21(2) & 4.232 & 270(20) & 0.0004(2)\\
Pu-Ga(2) & 24
&  0.0038(4) &  5.16(3) & 5.165 & 270(20) & 0.0012(4)\\
\\
Co-Ga(2) & 8
&  0.0019(1) &  2.45(1) & 2.471 & 250(20) & 0.0004(2)\\
Co-Pu & 2
&  0.005(3) &  3.38(5) & 3.393 & & \\
Co-Co & 4
&  0.007(7) &  4.18(8) & 4.232 &  & \\
Co-Ga(1) & 8
&  0.006(1) &  4.5(1) & 4.524 & & \\
Co-Ga(2) & 16
&  0.0030(6) &  4.84(2) & 4.901 & & \\
\\
Ga(1)-Pu & 4 & 0.0022(2)\footnotemark[1] & 2.97(1)\footnotemark[2]\footnotemark[7] & 2.993\\
Ga(1)-Ga(2) & 8 & 0.0053(2)\footnotemark[3]  & 2.95(1)\footnotemark[4]  & 2.993\\
Ga(1)-Ga(1) & 8 & 0.0051(2)\footnotemark[5]  & 4.20(2)\footnotemark[6]\footnotemark[7]\footnotemark[8]  & 4.232\\
Ga(1)-Co & 8 & 0.0023(5)  & 4.51(1)  & 4.524\\
Ga(2)-Co & 2 & 0.017(2) & 2.43(1)\footnotemark[8] & 2.471\\
Ga(2)-Ga(2) & 1 & 0.0026(2)   & 2.45(1)\footnotemark[8] & 2.552\\
Ga(2)-Ga(2) & 4 & 0.0053(2)\footnotemark[3]  & 2.97(1)\footnotemark[2] & 2.993\\Ga(2)-Ga(1) & 2 & 0.0053(2)\footnotemark[3]  & 2.95(1)\footnotemark[4] & 2.993\\Ga(2)-Pu & 8 & 0.0022(2)\footnotemark[1] & 2.95(1)\footnotemark[4] & 2.993\\
Ga(2)-Ga(2) & 4 & 0.0051(2)\footnotemark[5] & 3.94(2) & 3.933\\
Ga(2)-Ga(2) & 5 & 0.0051(2)\footnotemark[5] & 4.20(2)\footnotemark[6] & 4.232\\
\end{tabular}
\end{ruledtabular}
\footnotetext[1]{$^{-f}$like symbols held equal}
\footnotetext[7]{$r_\textrm{Ga(1)-Ga(1)}=\sqrt{2}r_\textrm{Ga(1)-Pu}$}
\footnotetext[8]{$r_\textrm{Ga(2)-Ga(2)}=\sqrt{4r_\textrm{Ga(2)-Co}^2-r_\textrm{Ga(1)-Ga(1)}^2}$}
\label{fit_tbl}
\end{table*}
%*****************************************************************************

%*****************************************************************************
%*****************************************************************************
\section{Analysis and Results}
\label{Results}

Figure \ref{ks_fig} shows an example of the normalized oscillations in the absorption above each measured edge as a function of $k$, the photoelectron wave vector, for the fresh (0.2$\times 10^{-6}$~$\alpha$-decays per atom) and 3 year old (14.5$\times 10^{-6}$~$\alpha$-decays per atom) samples. A Fourier transform (FT) of such data (Fig. \ref{rs_fig})
produces peaks in the amplitude as a function of the distance $r$ from the absorbing atomic species corresponding to neighboring atoms. For instance, the dominant peak in the Pu edge data (Fig. \ref{rs_fig}a) corresponds to the 12 Pu-Ga neighbors at $\sim$3.0 \AA.
Note that the atomic scattering functions generate complicated lineshapes, causing shifts in the peak positions from the actual structure that are well
reproduced by calculations using FEFF7.\cite{FEFF7} Information about the local
structure is therefore obtained by fitting these calculated scattering functions
to these data.  The results of fits to data from fresh material that has not undergone a significant amount of $\alpha$-decay are reported in Fig. \ref{rs_fig} 
and Table \ref{fit_tbl}. The fit quality is excellent, and all the measured 
pair distances agree well with diffraction results. We note that the local 
environment around Ga is more complicated than around Pu or Co, and therefore 
constraints were necessary to reduce the number of fit parameters while still 
obtaining high quality fits. Unfortunately, constraints add an unknown amount 
of systematic error that is not reflected in the estimated errors, and we 
ascribe discrepancies between diffraction and EXAFS results from the Ga $K$-edge
data to this source. Where the data are of sufficient quality, the temperature
dependence of the mean-squared displacements of the pair distances, 
$\sigma^2$'s, were obtained and are well described by a 
correlated-Debye model\cite{Beni76} with reasonable values of the 
correlated-Debye temperatures, $\Theta_\textrm{cD}$'s, (for comparison to 
$\delta$-Pu, see Ref. \onlinecite{Nelson03}), and no
evidence of static disorder from the fitted offsets, $\sigma_\textrm{stat}^2$'s.  

\begin{figure}
\includegraphics[width=3.5in]{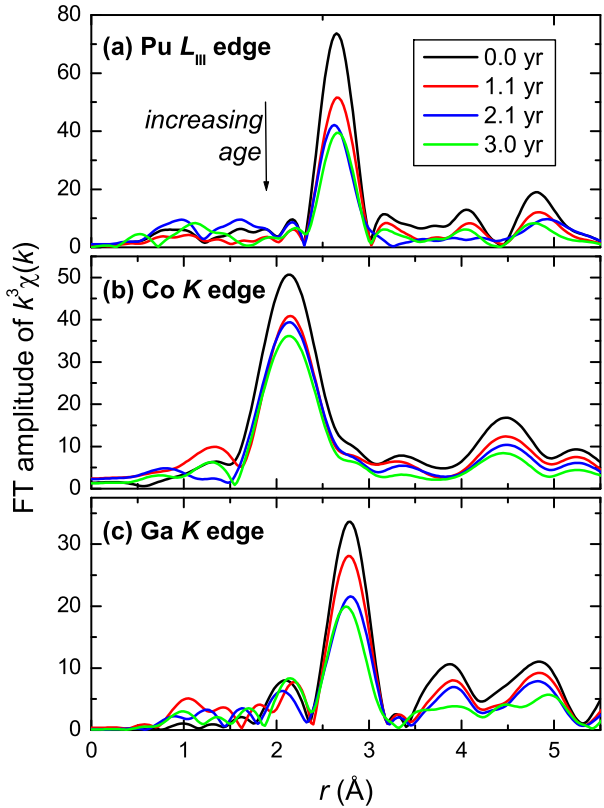}
\caption{(Color online) 
FT of the $k^3\chi(k)$ EXAFS data from the
(a) Pu $L_\textrm{III}$, (b) Co $K$, and the (c) Ga $K$ edges are shown 
for sample ages about one year apart. Samples have accumulated doses of 0.2$\times 10^{-6}$~$\alpha$-decays per atom (---), 5.4$\times 10^{-6}$~$\alpha$-decays per atom (\textcolor{red}{---}),  10.2$\times 10^{-6}$~$\alpha$-decays per atom (\textcolor{blue}{---}), and 14.5$\times 10^{-6}$~$\alpha$-decays per atom (\textcolor{green}{---}). Transform ranges are 
between 2.5-16.0 \AA$^{-1}$, 2.5-10.0 \AA$^{-1}$, and 2.5-14.5 \AA$^{-1}$, 
respectively, all Gaussian broadened by 0.3 \AA$^{-1}$.}
\label{rsall_fig}
\end{figure}

\begin{figure}
\includegraphics[width=3.5in,trim=0 0 0 0]{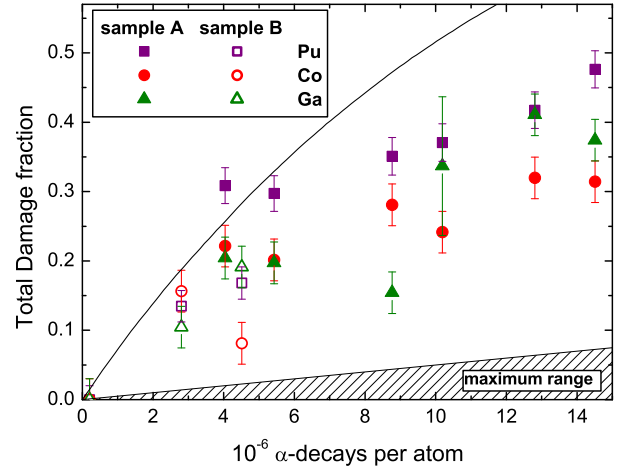}
\caption{(Color online) 
The total fraction of displaced atoms as a function of the
number of $\alpha$-decays per atom, as seen from 
each atomic species.  
Error bars are based on reproducibility. Absolute errors are about $\pm$5\%.
The line shows percolation behavior based on the
first two aged Pu data points for sample {\bf A}, and the hashed area gives the theoretical range of damage up to the amount expected in the absence of defect migration.
}
\label{dam_fig}
\end{figure}

Significant radiation damage effects are readily apparent in the raw data with 
a marked decrease in the overall amplitude of the spectra as samples are aged 
up to 3 years (Fig. \ref{rsall_fig}).  Preliminary fits to the data from the 
aged samples showed that the decrease in amplitude in the data is due both to 
a decrease in the overall scale factor $S_0^2$ and an increase in each atom 
pair's $\sigma^2$ with age. This situation indicates that there are at least 
three distinct regions within the aged samples: virtually undamaged, strongly 
damaged where the distance widths $\sigma_\textrm{s}^2$ are large enough that 
the local structure no longer contributes to the EXAFS amplitude, and mildly 
damaged where the $\sigma_\textrm{m}^2$'s only allow for a weak contribution. 
The latter regions may exist, for instance, on the edges of strongly damaged 
regions.  We therefore describe $f_\textrm{tot}$ as due to the sum of the 
strongly damaged fraction $f_\textrm{s}$ and the mildly damaged fraction 
$f_\textrm{m}$.  Since EXAFS amplitudes $A\sim1/\sigma$, as long as 
$\sigma_\textrm{m}$ is large enough, 
$f_\textrm{tot}\approx 1- S_0^{2\prime}(t)/S_0^2(0)$, where $S_0^{2\prime}(t)$ 
is obtained from fits where $\sigma^2(t)$ are fixed at $\sigma^2(0)$. That is, 
the total damage fraction in aged samples can be estimated by fixing 
most fitting parameters to those obtained from fresh-sample fits, and 
estimating $f_\textrm{tot}$ by the change in amplitude as given by $S_0^2$. 
These damage fractions are shown in Fig. \ref{dam_fig} and demonstrate strong 
damage production at apparently different rates for each atomic species.

%*****************************************************************************
%*****************************************************************************

Using this estimate of $f_\textrm{tot}$, $f_\textrm{s}$ is approximated by performing a fit where the $\sigma^2(t)$'s are
no longer constrained, in which case 
$f_\textrm{s}\approx(S_0^2(t)-S_0^{2\prime}(t))/S_0^2(0)$.
Correlations between the $S_0^2$ and $\sigma^2$ parameters are more difficult 
to control in such a procedure, but fits to Pu edge data indicate that between 60-80\% of the damage is due to the strongly damaged regions and
$\sigma_\textrm{m}$ is enhanced by $\sim$0.010-0.015~\AA$^2$ over the well-ordered regions for all of the aged samples measured.
Similar results are obtained from the Co and Ga edge data.

%*****************************************************************************
%*****************************************************************************

%*****************************************************************************
%*****************************************************************************
\section{Discussion}
\label{Discussion}

The accumulated damage after one year in sample {\bf A} is measured as between 20-30\% from the point of view of each of the constituent atoms in PuCoGa$_5$. This value is an order of magnitude higher than theoretical
estimates that only account for defects, or closer to two orders of magnitude if one allows for defect migration
after defect formation.\cite{Rubia99} It is important to note at this point that the damage enhancement, while possibly indicating more defects than expected, may also indicate significantly more defect-induced lattice strain and 
distortions.  In fact, this amount of damage is so large that, assuming a 
$\sim$10 nm damage cascade as expected for $\delta$-Pu, \textit{every} atom in 
the cascade is displaced from its equilibrium position.  Modern molecular 
dynamics calculations should, in principle, generate significant lattice
distortions around defects induced by radiation damage, but this effect has not,
to our knowledge, been reported in the literature for intermetallics.  In 
addition, the observed damage does not proceed at as fast a rate as 
extrapolated from lower accumulated doses. This disagreement is very
likely due to self-annealing\cite{Fluss04} caused by the 
room-temperature storage of the samples between measurements. Self-annealing is also likely the cause of the differences between the samples (sample {\bf B} appears to be damaging at a somewhat slower rate), as the exact history of the storage conditions for each sample then becomes important. In addition, differences between the constituent atomic species indicate different defect production or migration rates. These issues should be studied in other materials to further explore the role of self-annealing and atomic-species effects.

Another way of 
describing radiation damage is to consider that a decay event generates enough
heat within the damage cascade that the material locally melts and then
rapidly resolidifies, thus quenching disorder into the cascade region
from the high temperature state
and potentially also creating a distribution of competing structural phases.
This picture is similar to the ``thermal spike'' model as originally proposed
by Seitz,\cite{Seitz49} although it proved to be calculationally 
intractable.\cite{Kinchin55}
Applying molecular dynamics after (or during) Frenkel defect production seems
to combine the relevant aspects of these two schemes, and could, in principle,
generate the sort of distortions measured here in PuCoGa$_5$. Simulated EXAFS
data could be calculated from the results of future molecular 
dynamics calculations on PuCoGa$_5$ for a direct comparison of the efficacy of
such models.

These data largely explain the fast reduction of $T_c$ in PuCoGa$_5$, with 
$T_c/T_{c0} \approx 50$\% when $f_\textrm{tot} \approx 40\%$. This value is now 
greater than the lower-limit damage fraction of 
$f_\textrm{BdG}(T_c/T_{c0}=50\%) \approx$15\%,\cite{Franz97} as expected since 
not all damaged regions will create strong scattering.  In addition, it is 
likely that defects outside the superconducting planes scatter more weakly 
than defects on the in-plane sites.\cite{Petrovic02,Daniel05b} Decreasing the 
effective scattering strength would increase the necessary impurity densities 
in the theory.

Also shown in Fig. \ref{dam_fig} is the prediction of a cubic percolation model.
The time axis is chosen such that the model 
agrees with the first aged Pu edge data points for sample {\bf A}. Because of the strong damage that occurs within a cascade, the superconducting 
fraction is likely more closely related to the fraction of the 
material that exists within the volume between the edges of damage cascades, 
which is better described by the percolation model depicted in 
Fig. \ref{dam_fig}.  According to this extrapolation one might expect 
superconductivity to cease between 3.5 and 4 years for these samples 
($f_\textrm{tot}\sim0.8$), roughly consistent with the data in 
Fig. \ref{Tc_fig}, although no samples have yet been observed to become 
non-superconducting at this time. This
simplification, of course, doesn't account for any proximity effects, which 
should increase this time period, or any increased impurity scattering, which 
would decrease this time.

The emerging physical picture of PuCoGa$_5$ is one where the recoiling U nucleus 
generates much more damage than expected based on models of
elemental Pu.  This damage is likely dominated by near-neighbor lattice 
distortions that extend into the second coordination sphere or beyond,
possibly generating local distributions of impurity phases. This damage
is so severe that it encompasses all the atoms in a given damage cascade. The
effective damage rate is slowed as the material anneals at room temperature. 
Photoemission results indicate that the 5$f$ electrons have both local and itinerant character in PuCoGa$_5$.\cite{Joyce03} The partially localized $f$ electrons in the well-ordered material are likely further localized in the damaged regions, in analogy to $\delta$-Pu.\cite{McCall06} This view is consistent with x-ray absorption near-edge measurements on aged PuCoGa$_5$ samples.\cite{Booth06} Radiation damage therefore probably creates non-superconducting material both due to enhanced localization and strong defect scattering in these regions, although a proximity
effect could still allow some superconductivity. Annealed 
areas within a damage cascade probably would not be superconducting 
due to their limited size, unless they reach the edge of a cascade.

These data support and extend the conclusion of 
Farnan \textit{et al.}\cite{Farnan07,Farnan01}
that radiation damage occurs at a much faster rate than current theoretical 
predictions. In their work, the measured damage production rate is about 5 times
higher than the theoretical prediction for the unrelaxed defect production rate
in zircons. In the present work, we find a production rate at least 10 times faster
than the prediction in an intermetallic, and observe
deviations from a percolation model that we ascribe to annealing effects. The
role of annealing should be carefully considered in studies of zircons and related 
potential nuclear-waste storage materials.

%*****************************************************************************
%*****************************************************************************
\section{Conclusion}
\label{Conclusion}

Local structure data on samples of PuCoGa$_5$ demonstrate a well-ordered local lattice structure that agrees with the long-range average structure obtained by diffraction measurements. After the sample has aged long enough to accumulate a significant total number of $\alpha$-decays, the local structure exhibits strong disorder, primarily through a reduction in the amplitude of the EXAFS oscillations, but also in the pair-distance distribution variances $\sigma^2$'s. This disorder affects between 20-30\% of sample {\bf A} after one year ($\sim4\times10^{-6}$~$\alpha$-decays per atom), followed by a somewhat slower damage accumulation rate. Theoretical estimates that only account for defects predict at least an 
order of magnitude less damage, not including the damage-reducing factors of 
self-annealing and defect migration. These data help explain the fast reduction 
of the superconducting transition temperature both in terms of defect 
scattering and a simple percolation model.

These results underscore the need for more local structure studies of radiation damage in general 
and especially in PuCoGa$_5$. Only through better theoretical models and 
atomic-level probes can we understand the detailed 
electronic and structural properties of damaged regions and how they couple 
to superconductivity. In particular, direct comparisons between damage
cascade structures and local structure measurements should be pursued.
Such studies will have ramifications
not only for understanding superconductivity in 
PuCoGa$_5$, but also the unusual properties of $\delta$-Pu, and the field of radiation damage in general.

%*****************************************************************************
%*****************************************************************************
\section*{Acknowledgments}

We thank M. Fluss, M. Graf, A. Kubota, L. Soderholm, J. Thompson and W. Wolfer for enlightening discussions and 
W.-J. Hu for assistance in loading one of the plutonium samples.  Supported
by the 
U.S. Department of Energy (DOE) under Contract No. DE-AC02-05CH11231.
X-ray absorption data were collected at the Stanford Synchrotron Radiation 
Laboratory, a national user facility operated by Stanford University on
behalf of the DOE/OBES. Work at Los Alamos was performed under the auspices of the U. S. DOE.

\bibliography{../bib/bibli}

\end{document}